\documentclass[12pt,aps]{revtex4}
\usepackage[cp1251]{inputenc}

\input{epsf}
\begin{document}
\author{G.G.Kozlov}
\title{ Calculation of Anderson localization criterium for a one dimensional chain with diagonal disorder}

\begin{abstract}
For a one dimensional half-infinite chain with diagonal disorder we calculated
the ultimate at $t\rightarrow\infty$ value of the average excitation
density at the edge site $D$ if at $t=0$ the
excitation was localised at the edge site (Anderson' s creterium). We obtained
the following results: i) for the binary disordered chain we derived the close
expression for $D$ which is exact in the limit of low concentration of defects and
is valid for an arbitrary energy of defects $\varepsilon$. In this case $D$ demonstrated the non
analytical dependence on $\varepsilon$. ii) The close expression for $D$ is obtained for the
case of an arbitrary small disorder. iii) The relative contribution of states
with specified energy to $D$ is calculated. All the results obtained are in complete
agreement with computer simulation.
\end{abstract}
\maketitle
\section{Introduction, the problem setting, and main results }

The mathematical models of contemporary physics of disordered systems
 can be divided into two classes -- continuous
 and discrete. The continuous models are those in which
 the {\it Shredinger  equation}  $(-\Delta+v(r))\psi=E\psi$ with
  random potential  $v(r)$ is studied
 while the discrete models are dealing with  the {\it random matrix}
 of  Hamiltonian. These two types of models have much in
  common but despite the similarity they may require essentially  different methods
 of analysis due to the following reason.
 It is well known  \cite{Lif} that the possibility of the {\it localized} states whose wave functions
 are essentially differes from zero within the finite region when the system volume
 runs to infinity is one of the most important properties of the homogeneous
 disordered systems. For the continuous models of
 infinite volume the localized states can be recognized by dividing
 the energy spectrum into discrete and continuous parts with square
 integrated states of the discrete part considered to be localized.
 The energy spectrum of discrete models being the spectrum of random
 matrix is always discrete and for this reason one should use another
 criterium for states  characterization in this case.
 The Anderson criterium is one of the relevant ones\cite{And,Lif}.

 The deep theoretical analysis is possible for the one dimensional models of
disordered systems which we  have in mind hereafter.
 It is commonly accepted
that the spectrum of  one dimensional Shredinger equation with random potential is
completely discrete what is correspond to localization of all states for an
arbitrary low disorder \cite{Lif}. The validity of this statement to the discrete models is
arguable because of completely discrete character of  spectrum of such models
even in the absence of disorder when all states are delocalized. In present
paper we study the character of states in the sense of Anderson criterium for
the case of  one dimensional chain (discrete model) with diagonal disorder. The simple
physical sense of this criterium allows one to apply it both to discrete and
continuous models.

Let's turn to the problem setting and consider the one dimensional
diagonally disordered discrete model with the Hamiltonian whose
matrix has the following entries:

\begin{equation}
 H_{r,r'}=\delta_{r,r'}\varepsilon_r+ \delta_{r,r'+1}+\delta_{r,r'-1},
 \hskip10mm
 r,r'=1,...,N
 \end{equation}

Such Hamiltonian describe the Frenkel exciton in the chain
 consisting of N two-level atoms in the nearest neighbour
approximation. In this model the level splitting $\varepsilon_r$ of r-th atom
 is considered to be the random variable distributed in accordance
 with the function $P$ which is supposed known.

The unit non diagonal elements of Eq.(1) are specify the scale of energy.
Everywhere  below the thermodynamic limit $N\rightarrow\infty$ is implied. For this
model we now set the following problem. Let us suppose that the edge atom was
  prepared in the excited  state at $t=0$  and one should calculate the probability $D$ for this
atom to remain in  the excited state at $t\rightarrow\infty$.
 From the mathematical point of view it means that the initial state
 of the system is described by the wavefunction (column-vector)
  ${\bf\Psi}(0)$ with entries $\Psi_r(0)=\delta_{r,N}$
 and one should find $D=\langle|\Psi_N(t\rightarrow\infty)|^2\rangle$
 where angular brackets denote
the averaging  over the  random
 level splittings $\varepsilon_r$.
  The temporary dependance of the wavefunction can be written as:
 ${\bf \Psi}(t)=\exp\bigg(\imath {\bf H}t\bigg){\bf \Psi}(0)$.
Consequently the quantity of interest $D$ can be expressed in
terms of  eigen vectors
 ${\bf \Psi}^\lambda$
and eigen numbers  $E_\lambda, \lambda=1,...,N$ of the matrix Eq.(1) as:

\begin{equation}
D=\langle|\Psi_N(t\rightarrow\infty)|^2\rangle=\lim_{t\rightarrow\infty}\bigg\langle
\sum_{\lambda\lambda'}|\Psi^\lambda_N|^2|\Psi^{\lambda'}_N|^2\exp\imath
(E_\lambda-E_{\lambda'})t\bigg\rangle=
\bigg\langle\sum_\lambda |\Psi^\lambda_N|^4\bigg\rangle
\end{equation}

The similar quantities were analysed in the numerical study of disordered chains
for instance in \cite{Mal}.  Regarding  $D$
 one can make the following
 qualitative conclusions. Suppose all the  eigen functions of
Hamiltonian Eq.(1) are delocalized in the sense of their amplitude
 being nearly the same within an arbitrary region of the chain. The
 amplitude squared of such functions at the edge site can be
estimated as $|\Psi_N^\lambda|^2\sim 1/N$ with all $N$
 eigen functions giving  nearly the
same contribution $|\Psi_N^\lambda|^4\sim 1/N^2$ to Eq.(2).
 Consequently, if the states of Eq.(1)
are delocalized in the above sense then in the thermodynamic
limit $D\sim 1/N\rightarrow 0$.
 Let's consider now the situation when some of the eigen
 functions of Eq.(1) are localized  in the sense of their
amplitude being essentially non-zero only in some restricted
 region of the chain with the size of this region being
independent on $N$ when $N\rightarrow\infty$.
 Being defined only by the functions whose amplitude is essentially
non-zero at the edge site the contribution of functions of this
type to Eq.(2) will not depend on $N$.  In this case $D$ remain
finite in the thermodynamic limit. Having in mind all   above
qualitative conclusions one can introduce Anderson's criterium as:
{\it If $D$ is remain finite in the thermodynamic limit
 then in the set of eigen functions of
 Hamiltonian Eq.(1) there exist ones localized in the sense of Anderson's criterium.}

 To get some information concerning the degree of localization of
 eigen vectors of Eq.(1) in  some (specified) spectral range  we now introduce
the "participation" function defined as:

\begin{equation}
W(U)dU=\bigg\langle\sum_{E_\lambda\in[U,U+dU]} |\Psi^\lambda_N|^4\bigg\rangle,
\end{equation}

Obviously $D=\int W(U)dU$. Reasoning analogous to that presented  above shows that if
 all the states within the interval  $[U,U+dU]$ are delocalized then $W(U)=0$.
In the opposite case $W(U)$  differs from zero.
 Moreover, the function (3) provide quantitative information concerning
 the average value of eigen vectors of  random matrix Eq.(1) at
the edge site within the spectral interval $[U,U+dU]$.
The fantastic capabilities of modern personal computers
allows one to perform direct diagonalisation of  the matrix Eq.(1) for $N\sim 1000$  and more
 and in such numerical experiment to "observe" quantities Eq.(2) and Eq.(3).
 The theoretical
calculation of these quantities is the main  goal of this paper.

 The main results obtained in the present paper are:

1. The perturbation theory for calculation of joint distribution
 function of advanced and retarded Green's functions of
 Hamiltonian Eq.(1) is developed.

2. For the binary disordered one dimensional system,
described by  Hamiltonian (1) with the atomic
splittings $\varepsilon_r$ being equal to zero with
probability $1-c$ and  equal to $\varepsilon$ with
probability $c$ ($0<c<1$) the following expressions
 for $D$ and $W(U)$ function are obtained:
\begin{equation}
D={ c\over 4\pi}\int_{-2}^2
 dU(4-U^2)^{3/2}\hskip1mm\hbox{ln}\bigg({\varepsilon^2\over 4-U^2}+1\bigg)+
c\hskip1mm\Theta(|\varepsilon|-1) \bigg({\varepsilon^2-1\over\varepsilon^2}\bigg)^2+
O(c^2)
\end{equation}
\begin{equation}
W(U)={ c\over 4\pi}\Theta(2-|U|)
 (4-U^2)^{3/2}\hskip1mm\hbox{ln}\bigg({\varepsilon^2\over 4-U^2}+1\bigg)+
\end{equation}
$$
+c\Theta(|\varepsilon|-1)\bigg({\varepsilon^2-1\over\varepsilon^2}\bigg)^2
\delta\bigg(U- {\varepsilon^2+1\over\varepsilon}\bigg)
+O(c^2)
$$
 Nonanalyticity with respect to $\varepsilon$ is related to the
occurrence (when $|\varepsilon|>1$) of the edge state  with eigen energy
 $U_0=(\varepsilon^2+1)/\varepsilon$.

3. For the wide class of disordered systems described by  Hamiltonian
 (1) with random atomic splittings $\varepsilon_r$ having  the
 distribution functions in the form $P_\Delta(x)=p(x/\Delta)/\Delta$ (where
 $ p(x)>0,\int p(x)dx=1,\int p(x)xdx=0,\int p(x)x^2dx=M_2$)
 the following expressions for $D$ and $W(U)$ function are obtained:
 \begin{equation}
 D={\Delta^2 M_2\over 2}+O(\Delta^3),\hskip10mm
 W(U)=\Theta(2-|U|){\Delta^2 M_2\over 4\pi}\sqrt{4-U^2}+O(\Delta^3)
 \end{equation}

All the results obtained in the paper were verified by the computer
 simulation which shows that when $D<0.1$ the deviation of
formulas Eq.(4 -- 6) from the  numerical results is less than
$10\%$. In this case the degree of disorder may be rather large.
For example formula Eq.(6) is still valid when $\varepsilon_r$ values
 are uniformly distributed within the interval $[-0.25,0.25]$.

\section{ Statistics of Green's functions}

It is easy to show that the quantity  $\langle|\Psi_N(t)|^2\rangle$
   (with the quantity of interest  $D=\langle|\Psi_N(\infty)|^2\rangle$)
  can be calculated as:

\begin{equation}
\langle|\Psi_N(t)|^2\rangle=\lim_{V_{1,2}\rightarrow +0}{1\over 4
\pi^2}\int dU_1dU_2\exp[\imath (U_1-U_2)t]\langle\gamma(U_1-\imath V_1)\gamma(U_2+\imath
V_2)\rangle
\end{equation}

 Where $\gamma(\Omega)$ -- is the edge Green's function (EGF)
 for Hamiltonian Eq.(1):

\begin{equation}
\gamma(\Omega)\equiv\sum_\lambda {|\Psi_N^\lambda|^2\over \Omega-E_\lambda}
\end{equation}

To calculate the mean  product of two Green's functions entering  Eq.(7)
 one should know their joint distribution function.
 We obtain the equation for this function by generalising the Dyson's method
    \cite{Dyson,Lif}.
 Let's denote  by $\gamma(\Omega_i), i=1,2$ the edge Green's functions Eq.(8)
 with complex energies   $\Omega_1\equiv U_1-\imath V_1$ and
  $\Omega_2\equiv U_2+\imath V_2$ and add  one more atom with
 splitting $\varepsilon$ to the chain. Then as it is shown in \cite{Dyson,Lif}
EGF of  the chain with added atom  $\tilde{\gamma}(\Omega)$
 can be expressed in terms of EGF of the initial chain as:

   \begin{equation}
\tilde{\gamma}(\Omega)={1\over \Omega-\varepsilon-\gamma(\Omega)}
   \end{equation}

To describe the EGFs of the initial chain  $\gamma(\Omega_i), i=1,2$
 we introduce the distribution function $\eta$ defined
 in such a way that  the quantity
 $\eta(x_1,y_1,x_2,y_2)dx_1dy_1dx_2dy_2$ gives the probability of
  Re\hskip1mm$\gamma(\Omega_i)\in[x_i,x_i+dx_i]$ and
 Im\hskip1mm$\gamma(\Omega_i)\in[y_i,y_i+dy_i], i=1,2$.
  We denote  by $\tilde{\eta}$ the analagous function for the
 chain with added atom.
  The relation Eq.(9) allows one to express  $\tilde{\eta}$
 in terms of   $\eta$ and   the distribution function of  atomic splittings
 $P(\varepsilon)$:

\begin{equation}
\tilde{\eta}(\tilde{x_1},\tilde{y_1},\tilde{x_2},\tilde{y_2})=\int\delta\bigg( \tilde{x_1}
-\hbox{Re}{1\over\Omega_1-\varepsilon-x_1-\imath y_1}\bigg)
 \delta\bigg( \tilde{y_1}
-\hbox{Im}{1\over\Omega_1-\varepsilon-x_1-\imath y_1}\bigg)\times
\end{equation}
$$
\delta\bigg( \tilde{x_2}
-\hbox{Re}{1\over\Omega_2-\varepsilon-x_2-\imath y_2}\bigg)
 \delta\bigg( \tilde{y_2}
-\hbox{Im}{1\over\Omega_2-\varepsilon-x_2-\imath y_2}\bigg)
\eta(x_1y_1x_2y_2)P(\varepsilon) dx_1dy_1dx_2dy_2d\varepsilon
$$
%\newpage

In the thermodynamic limit $N\rightarrow\infty$
 it must be $\eta=\tilde{\eta}$.
 Calculating the integrals with $\delta$-functions in Eq.(10)
 we obtain for the steady state function $\eta$ the following equation:

 \begin{equation}
(x^2_1+y^2_1)^2(x^2_2+y^2_2)^2\eta(x_1,y_1,x_2,y_2)=
\end{equation}
$$
=\int\eta\bigg(
U_1-\varepsilon-{x_1\over x^2_1+y^2_1},\hskip2mm -V_1
+\hskip1mm{y_1\over x^2_1+y^2_1},\hskip2mm
U_2-\varepsilon-{x_2\over x^2_2+y^2_2},\hskip2mm V_2
+\hskip1mm{y_2\over x^2_2+y^2_2}
\bigg)P(\varepsilon) d\varepsilon
$$

Using the function $\eta$ one can introduce the mean  product of the
 advanced and retarded Green's functions entering Eq.(7) as a sum of
four terms:

\begin{equation}
\langle\gamma(\Omega_1)\gamma(\Omega_2)\rangle=\int\eta(x_1y_1x_2y_2)[x_1x_2-y_1y_2+\imath(y_1x_2+y_2x_1)]
dx_1dx_2dy_1dy_2\equiv
\end{equation}
$$
=\langle x_1x_2\rangle-\langle y_1y_2\rangle+\imath\langle y_1x_2\rangle+\imath\langle y_2x_1\rangle
$$

Using the expression Eq.(8) it is easy to see that these terms can be
 written in the representation of Hamiltonian Eq.(1)  as:

\begin{equation}
 \langle x_1x_2\rangle=\bigg\langle
 \sum_{\lambda\lambda'}{|\Psi_N^\lambda|^2|\Psi_N^{\lambda'}|^2 (U_1-E_\lambda)(U_2-E_{\lambda'})\over
[(U_1-E_\lambda)^2+V_1^2][(U_2-E_{\lambda'})^2+V_2^2]}\bigg\rangle
\end{equation}
$$
 -\langle y_1y_2\rangle=
\bigg\langle
\sum_{\lambda\lambda'}{|\Psi_N^\lambda|^2|\Psi_N^{\lambda'}|^2V_1V_2\over
[(U_1-E_\lambda)^2+V_1^2][(U_2-E_{\lambda'})^2+V_2^2]}\bigg\rangle
$$
$$
\imath\langle y_1x_2\rangle=\imath\bigg\langle
\sum_{\lambda\lambda'}{|\Psi_N^\lambda|^2|\Psi_N^{\lambda'}|^2V_1(U_2-E_{\lambda'})\over
[(U_1-E_\lambda)^2+V_1^2][(U_2-E_{\lambda'})^2+V_2^2]} \bigg\rangle
$$
$$
\imath\langle x_1y_2\rangle=
-\imath\bigg\langle
\sum_{\lambda\lambda'}{|\Psi_N^\lambda|^2|\Psi_N^{\lambda'}|^2V_2(U_1-E_{\lambda})\over
[(U_1-E_\lambda)^2+V_1^2][(U_2-E_{\lambda'})^2+V_2^2]} \bigg\rangle
$$

Thus,  $\langle|\Psi_N(t)|^2\rangle$ Eq.(7) can be represented as a sum
 of four contributions:

\begin{equation}
\langle|\Psi_N(t)|^2\rangle=\Delta_{\langle x_1x_2\rangle}+\Delta_{\langle
y_1y_2\rangle}+\Delta_{\langle y_1x_2\rangle}+\Delta_{\langle x_1y_2\rangle}.
\end{equation}

We present the calculation of   $\Delta_{\langle y_2x_1\rangle}$
 as an example:

\begin{equation}
\Delta_{\langle y_2x_1\rangle}={\imath\over 4\pi^2}\lim_{V_{1,2}\rightarrow +0}
 \int dU_1dU_2\exp\imath(U_1-U_2)t\hskip2mm\langle y_2x_1\rangle=
\end{equation}
$$
=-{\imath\over 4\pi^2}\lim_{V_{1,2}\rightarrow +0}
 \int dU_1dU_2\hskip2mm\exp\imath(U_1-U_2)t \hskip2mm\bigg\langle\sum_{\lambda\lambda'}
{|\Psi_N^\lambda|^2|\Psi_N^{\lambda'}|^2V_2(U_1-E_\lambda)\over
[(U_1-E_\lambda)^2+V_1^2][(U_2-E_{\lambda'})^2+V_2^2]}\bigg\rangle=
$$
$$
={1\over 4}\bigg\langle\sum_{\lambda\lambda'}
|\Psi_N^\lambda|^2|\Psi_N^{\lambda'}|^2\exp\imath(E_\lambda-E_{\lambda'})t\bigg\rangle
$$

The similar calculations shows that all four contributions Eq.(14) to
$\langle|\Psi_N(t)|^2\rangle$
 are equal to each other:
  $\Delta_{\langle x_1x_2\rangle}=\Delta_{\langle
y_1y_2\rangle}=\Delta_{\langle y_1x_2\rangle}=
\Delta_{\langle x_1y_2\rangle}$ and therefore:

\begin{equation}
\langle|\Psi_N(t)|^2\rangle=4\Delta_{\langle y_2x_1\rangle}
\end{equation}

The following remark is important for what is hearafter.
  Suppose that the integration over $U_{1,2}$ in Eq.(15) runs over the
small region  $U_{1,2}\in [U,U+dU]$ only and we are interested in
 the behaviour of $ \Delta_{\langle y_2x_1\rangle}$ contribution at
 $t\rightarrow\infty$.
 In this case the sum in the last string of Eq.(15) will
 contain the states with energies $E_\lambda\in[U,U+dU]$ only.
  So, it is seen that such restriction of the integration region allows
 one to calculate the "participation" function Eq.(3).

 Thus, the problem reduced to solving of the Eq.(11) for the joint
 probability distribution function $\eta$.
  The fact that when calculating the contributions Eq.(13) the
 limit  $V_{1,2}\rightarrow +0$ is implied one can use
 for reducing the problem to studying the equation which is much easier
 than Eq.(11). To do this we note that if  $V_1=V_2=0$ then the solution of Eq.(11)
 can be presented in the form:

 \begin{equation}
\eta(x_1,y_1,x_2,y_2)\bigg|_{V_{1,2}=0}=\delta(y_1)\delta(y_2)\rho(x_1,x_2),
\end{equation}

where the depending on $U_{1,2}$ function $\rho(x_1,x_2)$
 satisfy the following equation:

\begin{equation}
x_1^2x_2^2\rho(x_1,x_2)=\int P(\varepsilon)
\rho(U_1-\varepsilon-1/x_1,\hskip2mm U_2-\varepsilon-1/x_2) d\varepsilon
\end{equation}

 Now let us perform the calculation of the quantity  $\langle y_2x_1\rangle$
 taking into account that for small $V_{1,2}$ the solution of Eq.(11)
 goes to Eq.(17).
 Using the fact that the function $\eta$ is satisfy to Eq.(11)
 one can write the following expression for the mean of
 interest   $\langle y_2x_1\rangle$:

 \begin{equation}
\langle y_2 x_1\rangle=\int\eta(x_1y_1x_2y_2) y_2x_1
dx_1dx_2dy_1dy_2=
\end{equation}
$$
\int {dx_1dx_2dy_1dy_2d\varepsilon \hskip0.5mm y_2 x_1\over (x_1^2+y_1^2)^2
(x_2^2+y_2^2)^2}\eta
\bigg(U_1-\varepsilon -{x_1\over x_1^2+y_1^2},-V_1+{y_1\over x_1^2+y_1^2},
U_2-\varepsilon -{x_2\over x_2^2+y_2^2},V_2+{y_2\over x_2^2+y_2^2}\bigg)P(\varepsilon)=
$$
By replacing the variables:

\begin{equation}
-{x_k\over y_k^2+x_k^2}\rightarrow x_k,\hskip10mm
{y_k\over y_k^2+x_k^2}\rightarrow y_k\hskip10mm k=1,2
\end{equation}
and calculating the corresponding  Jacobians one can continue the
 equality Eq.(19) as:
$$
=-\int dx_1dx_2dy_1dy_2d\varepsilon\hskip2mm\eta(U_1-\varepsilon +x_1,y_1-V_1,
U_2-\varepsilon +x_2,V_2+y_2){x_1\over y_1^2+x_1^2}{y_2\over y_2^2+x_2^2}
$$

Taking advantage of the fact that when $V_{1,2}\rightarrow +0$ the function $\eta$
 is close to Eq.(17) one can conclude  that in the region where the
expression under this integral is essentially differs from zero the
 following estimations are valid:  $y_1\approx V_1, y_2\approx -V_2$,
 with the accuracy of estimations increasing in the limit
 $V_{1,2}\rightarrow +0$.
 In this limit   $y_2/(y_2^2+x_2^2)\rightarrow
 -\pi\delta(x_2)$, and  $y_1\rightarrow +0$.
 Having all this in mind one can now perform the integration over $x_2$,
 replace the function $\eta$ by its ultimate expression Eq.(17) and
 finally get:

\begin{equation}
\langle y_2x_1\rangle=\pi\int d\varepsilon{dx\over x}\hskip2mm\rho(U_1-\varepsilon+x ,
U_2-\varepsilon)P(\varepsilon),
\end{equation}
Here  we imply the main value of the integral.
 The same calculations can be performed for
  $\langle y_1 x_2\rangle$, $\langle y_1y_2\rangle$
 and $\langle x_1x_2\rangle$.

 In the end one should take into account the following important remark.
It is seen from the Eq.(15) that if $\langle y_2x_1\rangle$
 has no singularity at $U_2-U_1\equiv\omega=0$  then
  $\lim_{t\rightarrow\infty}\Delta_{\langle y_2x_1\rangle}=0$.
 Therefore the non-zero value of $D$ is related to the
 occuarence of the singularity of $\langle y_2x_1\rangle$
  at  $\omega=0$.
 Thus, for  calculation of  $D$ it is sufficient to solve the Eq.(18)
 for small $\omega$, extract the singular part and use it for the calculation
 of  ultimate at  $t\rightarrow\infty$ behaviour of the integral in Eq.(15).
 This will be done in the next sections

\section{Binary disorder }

In the case of binary disorder mentioned in the first section the distribution
function of atomic levels splittings  has the form:

 \begin{equation}
 P(y)=(1-c)\delta(y)+c\delta(y-\varepsilon),\hskip10mm 0<c<1
 \end{equation}
 and Eq.(18) can be written as:
 \begin{equation}
 x_1^2 x_2^2\rho(x_1x_2)=(1-c)\rho(U_1-1/x_1,U_2-1/x_2)+
  c\rho(U_1-\varepsilon-1/x_1,U_2-\varepsilon-1/x_2)
 \end{equation}
We now introduce the function $\rho$ as a  series in powers of  $c$:
 \begin{equation}
 \rho=\sum_{n=0}^\infty c^n\rho_n
 \end{equation}
By substituting this series into  Eq.(23) and
 equating the coefficients at equal powers of $c$
 one can obtain:
\begin{equation}
c^0:\hskip10mm \rho_0(x_1x_2)x_1^2x_2^2=\rho_0(U_1-1/x_1,U_2-1/x_2)
\end{equation}
%\newpage
\begin{equation}
\hskip3mm c^1:\hskip10mm\rho_1(x_1x_2)x_1^2x_2^2=\rho_1(U_1-1/x_1,U_2-1/x_2)+
\end{equation}
$$
+\rho_0(U_1-\varepsilon-1/x_1,U_2-\varepsilon-1/x_2)-
\rho_0(U_1-1/x_1,U_2-1/x_2),\hskip3mm\hbox{and so on}
$$
To calculate the value of $D$ up to the terms of $\sim c^2$ it
is sufficient to calculate the mean Eq.(21) with the same accuracy.
 By substituting  Eq.(22) and Eq.(24) to Eq.(21) we obtain:
\begin{equation}
\langle y_2x_1\rangle=\pi\int{dx\over x}\bigg\{
(1-c)\rho_0(U_1+x,U_2)+c\rho_0(U_1-\varepsilon+x,U_2-\varepsilon)+
c\rho_1(U_1+x,U_2)\bigg\} + O(c^2)
\end{equation}
The first term in braces  gives the mean  $\langle y_2x_1\rangle$ (up to the factor $1-c$)
 for the  {\it completely ordered} chain (when $D=0$) and is not of
 interest for us.
 Thus, for the singular part of $\langle y_2x_1\rangle$ (we denote it as "sing")
 which we are interested in and for the ultimate at  $t\rightarrow\infty$ value
 of $\Delta_{\langle y_2x_1\rangle}$ we obtain the following expressions:
\begin{equation}
\hbox{sing}\langle y_2x_1\rangle=c\pi\int{dx\over x}\bigg\{
\rho_0(U_1-\varepsilon+x,U_2-\varepsilon)+
\rho_1(U_1+x,U_2)\bigg\} + O(c^2)
\end{equation}

$$
\lim_{t\rightarrow\infty}\Delta_{\langle y_2x_1\rangle}=
{\imath\over 4 \pi^2}\lim_{t\rightarrow\infty}
\int dU_1dU_2\exp\imath(U_1-U_2)t\hskip2mm
\hbox{sing}\langle y_2x_1\rangle
$$
To obtain the functions $\rho_0$  and $\rho_1$ entering Eq.(28) one should solve
the equations Eq.(25) and Eq.(26).
 We start the analysis of these equations from the most important case
when  $|U_{1,2}|<2$, i.e.  when energies of both Green's functions are
belong to the spectrum of bare Hamiltonian Eq.(1) at $\varepsilon_r=0$.

\subsection{The contribution of region  $|U_{1,2}|<2$}

By direct substitution one can see that in this case the exact solution of Eq.(25)
can be written in the close form:
\begin{equation}
\rho_0(x_1x_2)={\cal L}_{U_1}(x_1){\cal L}_{U_2}(x_2),
\end{equation}
where  ${\cal L}_U(x)$ is Lorentzian:
\begin{equation}
{\cal L}_U(x)={\sqrt{4-U^2}\over 2\pi}{1\over x^2-Ux+1}
\end{equation}
 Let us turn now to the Eq.(26) whose solution for $|U_{1,2}|<2$ we will construct
  using the system of special functions proposed  by the author in  \cite{Koz}.
  Below we briefly review the results obtained in \cite{Koz}.

 Define the depending on parameter  $U, |U|<2$ linear operator ${\cal H}_U$
 which acts on an arbitrary function in accordance with the following definition:
 \begin{equation}
 {\cal H}_U f(x)\equiv{1\over x^2} f(U-1/x)
 \end{equation}
 As it is shown in \cite{Koz} the eigen functions $\sigma^{(n)}_U(x)$ and eigen
numbers $\lambda_n$ (they can be numbered  by  integer $n$)  are defined by the
following relations:
\begin{equation}
\sigma_U^{(n)}(x)={\cal L}_U(x)\bigg[
{R^\ast-x\over R-x}\bigg]^n\equiv
{\cal L}_U(x)G^n(x),
\hskip3mm \lambda_n=\bigg({U+\imath\sqrt{4-U^2}\over
U-\imath\sqrt{4-U^2}}\bigg)^n,
\hskip3mm |\lambda_n|=1
\end{equation}
where
$$
R={U+\imath\sqrt{4-U^2}\over 2},\hskip10mm R^\ast={U-\imath\sqrt{4-U^2}\over
2},\hskip5mm R R^\ast=1
$$
 The map corresponding to operator ${\cal H}_U$  plays an important role in Eqs. (25), (26)
 and for this reason we will search for the solution of these equations in the
 form of the expansion into a set of  functions Eq.(32).
 To do this we use the rules of expansion of an arbitrary function  $f(x)$ in a
 set of functions Eq.(32) obtained in \cite{Koz}:
 \begin{equation}
 f(x)=\sum_{n=-\infty}^{+\infty}A_n\sigma^{(n)}_U(x)
 \end{equation}
 with coefficients  $A_n$ being defined by formulas:
 \begin{equation}
 A_n=\int {f(x)\over G^n(x)}dx
 \end{equation}
Using the functions Eq.(32) one can expand  $\rho_1(x_1x_2)$ in series:
\begin{equation}
\rho_1(x_1x_2)=\sum_{|n|+|m|\ne 0}C_{nm}\sigma_{U_1}^{(n)}(x_1)\sigma_{U_2}^{(m)}(x_2)
\end{equation}
Substituting the series Eq.(35) into Eq.(26) and making use of the properties of
functions Eq.(32) one can obtain:
\begin{equation}
\sum_{|n|+|m|\ne
0}C_{nm}\sigma^{(n)}_{U_1}(x_1)\sigma^{(m)}_{U_2}(x_2)[1-\lambda_n(U_1)\lambda_m(U_2)]=
\end{equation}
$$
={\rho_0(U_1-\varepsilon-1/x_1,U_2-\varepsilon-1/x_2)\over x_1^2 x_2^2}-{\cal
L}_{U_1}(x_1) {\cal L}_{U_2}(x_2)
$$
With the help of Eq.(33) and Eq.(34) one can expand the r.h.p. of this equation
 in a set  of functions Eq.(32) and obtain
the following expressions for the expansion  coefficients of $\rho_1$:
\begin{equation}
C_{nm}={J_n(U_1,\varepsilon) J_m(U_2,\varepsilon)\over 1-\lambda_n(U_1)\lambda_m(U_2)}
\end{equation}
where functions $J_n(U,\varepsilon)$ are  defined as:
\begin{equation}
J_n(U,\varepsilon)\equiv\int{{\cal L}_U(U-\varepsilon-1/x)\over G^n(x)}{dx\over
x^2}= \lambda_n(U)\int{{\cal L}_U(x-\varepsilon)\over G^n(x)}dx,
\end{equation}
The following (based on the properties of
$\sigma^{(n)}_U(x)$ functions) relations  are take place:
\begin{equation}
J_n(U,0)=0\hskip5mm\hbox{for}\hskip2mm n\ne 0,\hskip5mm
J_n(U,\varepsilon)=J_{-n}^\ast(U,\varepsilon)\Rightarrow
|J_n(U,\varepsilon)|^2=|J_{-n}^\ast(U,\varepsilon)|^2
\end{equation}
Thus, the solution of Eq.(26) for $\rho_1$ has the form:
\begin{equation}
\rho_1(x_1x_2)=\sum_{|n|+|m|\ne 0}
{J_n(U_1,\varepsilon)J_m(U_2,\varepsilon)\over 1-\lambda_n(U_1)\lambda_m(U_2)}
\sigma_{U_1}^{(n)}(x_1)\sigma_{U_2}^{(m)}(x_2)
\end{equation}
 As it was mentioned above only singular at $\omega=U_2-U_1\approx 0$ part of
 this expression is needed  for calculation of $D$.
 It easy to see that only terms with $m=-n$  of sum Eq.(40) possess  the required peculiarity
at  $\omega=0$.
Now we make the following replacing of symbols: $U_1\rightarrow U,
  U_2\rightarrow U+\omega$ and write down the expression for the denominators of
  these terms up to $\omega^2$:
\begin{equation}
1-\lambda_n(U)\lambda_{-n}(U+\omega)=-{2\imath n\omega\over\sqrt{4-U^2}}+O(\omega^2)
\end{equation}
%\newpage
Taking this into account we obtain the following expression for the singular
part of Eq.(40):
\begin{equation}
\hbox{sing}\hskip1mm\rho_1(x_1x_2)={\imath\sqrt{4-U^2}\over 2\omega}\sum_{n\ne 0}
{|J_n(U,\varepsilon)|^2\over n}\sigma_U^{(n)}(x_1)\sigma_U^{-n}(x_2)
\end{equation}
Let us turn now to  the Eq.(28).
 It is easy to see that the first item under  integral do not contribute to
the final result when $|U_{1,2}|<2$ due to the regularity of function $\rho_0$ Eq.(29) at $\omega=0$.
 For this reason only the second item (it depends on function  $\rho_1$) remain to be
 considered:
\begin{equation}
\hbox{sing}\langle y_2x_1\rangle=\hbox{sing}\hskip1mm
\pi c\int{dx\over x}\rho_1(U+x,U+\omega)=
\end{equation}
$$
=\imath\pi c {\sqrt{4-U^2}\over 2\omega}
\sum_{n\ne 0}{|J_n(U,\varepsilon)|^2\over n}
\sigma^{-n}_U(U)\int{dx\over x}\sigma_U^{(n)}(U+x)=
\imath c {4-U^2\over 4\omega}
\sum_{n\ne 0}{|J_n(U,\varepsilon)|^2\over n}\int dx x\sigma_U^{(n)}(x)
$$
 When evaluating the expression under integral we make the following replacing of
 variable  $x\rightarrow - 1/x$ and make use of the fact that the functions
  $\sigma^{(n)}_U(x)$ are eigen for the operator  ${\cal H}_U$ Eq.(31).
 Now we can calculate the ultimate at $t\rightarrow\infty$ behaviour of the
contribution Eq.(15):
\begin{equation}
\lim_{t\rightarrow\infty}\Delta_{\langle y_2x_1\rangle}\bigg |_{|U_{1,2}|<2}=
-{c\over 16\pi^2}
\lim_{t\rightarrow\infty}\int{\exp\imath\omega t\over \omega}d\omega\hskip2mm
\int_{-2}^2dU (4-U^2)
\sum_{n\ne 0}{|J_n(U,\varepsilon)|^2\over n}\int dx x\sigma_U^{(n)}(x)
\end{equation}
The ultimate behaviour at  $t\rightarrow\infty$ of the first integral do not
 depend on the region of integration over $\omega$:
\begin{equation}
\lim_{t\rightarrow\infty}\int{\exp\imath\omega t\over \omega}d\omega=
\int_{-\infty}^\infty {\exp\imath x\over x}dx=\imath\pi
\end{equation}
 The calculation of the first moments of functions $\sigma^{(n)}(x)$ and
the integrals Eq.(38) gives:
\begin{equation}
\int dx x\sigma_U^{(n)}(x)={\imath\over 2}{n\over |n|}\sqrt{4-U^2},
\hskip10mm
|J_n(U,\varepsilon)|^2=\bigg({\varepsilon^2\over \varepsilon^2+4-U^2}\bigg)^n
\end{equation}
Thus, Eq.(44) can be evaluated as:
\begin{equation}
\lim_{t\rightarrow\infty}\Delta_{\langle y_2x_1\rangle}\bigg |_{|U_{1,2}|<2}
 ={ c\over 16\pi}\int_{-2}^2
 dU(4-U^2)^{3/2}\sum_{n=1}^\infty{1\over n}{\bigg({\varepsilon^2\over \varepsilon^2+4-U^2}\bigg)^n}+ O(c^2)
 \end{equation}
And taking into account that
 $\sum_{n=1}^\infty q^n/ n=-\hbox{ln}(1-q)$,
 we finally obtain:
\begin{equation}
\lim_{t\rightarrow\infty}\Delta_{\langle y_2x_1\rangle}\bigg |_{|U_{1,2}|<2}
 ={ c\over 16\pi}\int_{-2}^2
 dU(4-U^2)^{3/2}\hskip1mm\hbox{ln}\bigg({\varepsilon^2\over 4-U^2}+1\bigg)+ O(c^2)
 \end{equation}

\subsection{The contribution of region $|U_{1,2}|>2$}

Let us return to the Eq.(28) and consider the contribution of the region   $|U_{1,2}|>2$
 to the integrals in Eq.(28).
 It is shown in the Appendix that the second item in braces in Eq.(28) (depending on $\rho_1$)
 do not contribute to the final result.
 Thus, the contribution of the region $|U_{1,2}|>2$ is defined by  the first (depending on
$\rho_0$) item of Eq.(28) only.
 As the expression for  EGF of the ordered chain is known
 \footnote { It can be obtained by solution of Eq.(9) at $\tilde\gamma=\gamma, \varepsilon=0$}
 the solution of Eq.(25) for the unperturbed joint probability distribution function
 $\rho_0$ (for   $|U_{1,2}|>2$) can be easily guessed:
\begin{equation}
\rho_0(x_1x_2)=\delta\bigg[
x_1-{U_1-\hbox{sign}(U_1)\sqrt{U_1^2-4}\over 2}
\bigg]
\delta\bigg[
x_2-{U_2-\hbox{sign}(U_2)\sqrt{U_2^2-4}\over 2}
\bigg]
\end{equation}
We calculate now the contribution of the first (depending on $\rho_0$)
item in braces (28) (we denote it by $I_1$):
$$
I_1={\imath c\over 4\pi}\int{dxdU_1dU_2\over x}\exp[\imath(U_1-U_2)t]\hskip2mm
 \rho_0(U_1-\varepsilon+x,U_2-\varepsilon)
$$
\begin{equation}
I_1={\imath c\over 4\pi}\int{dxdU_1dU_2\over x}\exp[\imath(U_1-U_2)t]\hskip2mm
\delta\bigg[
x-\varepsilon+{U_1+\hbox{sign}(U_1)\sqrt{U_1^2-4}\over 2}
\bigg] \times
\end{equation}
$$
\hskip30mm\times\delta\bigg[
{U_2+\hbox{sign}(U_2)\sqrt{U_2^2-4}\over 2}-\varepsilon
\bigg]
$$
The integrations over $U_{1,2}$ run over the regions $|U_{1,2}|>2$.
 Integrating Eq.(50) over $x$ we get:
\begin{equation}
I_1={\imath c\over 4\pi}\int{dU_1dU_2\exp\imath(U_1-U_2)t\over
\varepsilon-[U_1+\hbox{sign}(U_1)\sqrt{U_1^2-4}]/ 2
} \hskip2mm\delta\bigg[
{U_2+\hbox{sign}(U_2)\sqrt{U_2^2-4}\over 2}-\varepsilon
\bigg]
\end{equation}
 It is clear that the result do not depend on the sign of  $\varepsilon$
  and below we perform calculations for   $\varepsilon>0$.
 In this case $\delta$-function under the integral Eq.(51) gives zero for  $U_2<-2$
 and, moreover, if $0<\varepsilon<1$ this $\delta$-function is equal to zero identically.
 {\it Consequently the  contribution under consideration is differ from zero only when
  $|\varepsilon |>1$.}
 The function under integral in Eq.(51) has a $\delta$-peculiarity  with respect
to  $U_2$ and a pole-peculiarity with respect to  $U_1$
 To calculate these integrals (at $t\rightarrow \infty$) we take into account
the following property of the relevant function
 $\Phi(U)$ defined as:
\begin{equation}
\Phi(U)\equiv{U+\sqrt{U^2-4}\over 2}-\varepsilon
\end{equation}
This function is equal to zero at:
\begin{equation}
U=U_0\equiv{\varepsilon^2+1\over \varepsilon}
\end{equation}
and can be expanded at $U\approx U_0$ as:
\begin{equation}
\Phi(U)\approx (U-U_0){\varepsilon^2\over \varepsilon^2-1}
\end{equation}
Using this formula one can perform the integration over  $U_2$ in Eq.(51):
\begin{equation}
\int dU_2\exp[-\imath U_2t] \delta\bigg[
{U_2+\sqrt{U_2^2-4}\over 2}-\varepsilon
\bigg]={\varepsilon^2-1\over\varepsilon^2}\exp[-\imath U_0t]
\end{equation}
 Let us now turn to the integration over $U_1$ in Eq.(51).
 Due to the fact that one should calculate this integral at
$t\rightarrow\infty$ the contribution of pole-peculiarity is the only one of importance.
 Using the expansion Eq.(54) again one can obtain the following expression for this
integral:
\begin{equation}
\int_{U_1>2}{dU_1\exp[\imath U_1t]\over
\varepsilon-[U_1+\sqrt{U_1^2-4}]/ 2}\approx
-{\varepsilon^2-1\over\varepsilon^2}
\int_{-\infty}^{+\infty}{dU_1\exp[\imath U_1t]\over U_1-U_0}=
-\imath\pi\exp[\imath U_0t] \hskip1mm {\varepsilon^2-1\over\varepsilon^2}
\end{equation}
The approximate equality here becomes exact at  $t\rightarrow\infty$.
 Thus,
\begin{equation}
I_1=\Theta(|\varepsilon|-1){c\over 4}\bigg({\varepsilon^2-1\over\varepsilon^2}\bigg)^2
\end{equation}
The additional study shows that this contribution is related to the occurrence
(when $|\varepsilon|>1$) of the edge state with the energy $U_0$ defined by Eq.(53).
 To obtain the final formula for $D$ one should sum the contributions Eq.(57)
and Eq.(48) and in accordance with Eq.(16) multiply the result by four.
  Thus, we obtain the Eq.(4) for $D$ for the case of binary disordered chain.
 In accordance with the logic of the above calculation and taking into account
the remark after Eqs.(15,16) one can see that the "participation" function   $W(U)$
Eq.(3) is define by Eq.(5).

\section{An arbitrary small diagonal disorder }

 The analysis of the binary disordered chain presented above can be regarded as
 a consistent  perturbation theory for the statistics of advanced and retarded
 Green's functions with concentration $c$ of atoms with the level splitting  $\varepsilon$
 playing the role  of small parameter.
 The perturbation theory for the case of the chain with an arbitrary small
diagonal disorder can be constructed in a similar way. The relevant small
parameter can be defined as follows.

 Let the function $p(\varepsilon)$ possess the following properties:
$p(\varepsilon)>0$, $\int
p(\varepsilon)d\varepsilon=1$.
 With the help of this function we construct now the following family of the
atomic levels splittings   distribution functions  $P_\Delta(\varepsilon)$:
\begin{equation}
P_\Delta(\varepsilon)={1\over\Delta} p\bigg({\varepsilon\over \Delta}\bigg)
\end{equation}
 If the moments of $p(\varepsilon)$ function are:
\begin{equation}
M_n\equiv\int p(\varepsilon)\varepsilon^n d\varepsilon
\end{equation}
 then the moments of functions Eq.(58) can be expressed as:
 \begin{equation}
 \int P_\Delta(\varepsilon)\varepsilon^n d\varepsilon=\Delta^n M_n
 \end{equation}
 It is clear that one can consider $\Delta$ as a degree of disorder
  -- the system becomes ordered at  $\Delta\rightarrow 0$.
 Without loss of generality one can say that  $M_1=0$.
 Thus, one should construct the perturbation theory for the Eq.(18)
when  $P(\varepsilon)= P_\Delta(\varepsilon)$ with  $\Delta$ playing the role
of  small parameter.
 Therefore we now construct the expansion of  solution of Eq.(18) in powers
 of  $\Delta$ and we start from the case of $|U_{1,2}|<2$ which is of
 particular importance.
 For this reason we write the expansion of the function $\rho(x_1x_2)$ in the
vicinity of  $x_{10},x_{20}$:
\begin{equation}
\rho(x_1x_2)=\sum_{n,m=0}^\infty \rho_{nm}(x_1-x_{10})^n(x_2-x_{20})^m,\hskip10mm
\rho_{nm}\equiv{1\over n!
 m!}\hskip1mm{\partial^{n+m}\over\partial\theta_1^n\partial\theta_2^m}
 \hskip1mm\rho(\theta_1\theta_2)\bigg|_{\theta_i=x_{i0}}
\end{equation}
 By applying this expansion to the r.h.p of Eq.(18) with $x_{i0}=U_i-1/x_i, i=1,2$
 one can obtain the expansion of the function $\rho(U_1-\varepsilon -1/x_1,U_2-\varepsilon -1/x_2)$
 in powers of $\varepsilon$ and express the r.h.p. of Eq.(18) in terms of moments
 Eq.(59) and powers of  $\Delta$:
\begin{equation}
\int d\varepsilon P_\Delta(\varepsilon)\rho(U_1-\varepsilon-1/x_1,
 U_2-\varepsilon-1/x_2)= \sum_{n,m=0}^\infty
{(-\Delta)^{n+m}M_{n+m}\over n!
 m!}\hskip1mm{\partial^{n+m}\rho(\theta_1\theta_2)\over\partial\theta_1^n\partial\theta_2^m}
  \hskip1mm\bigg|_{\theta_1=U_1-1/x_1}^{\theta_2=U_2-1/x_2}
\end{equation}
 It follows from this that the l.h.p. of Eq.(18) (i.e. the function $\rho(x_1x_2)$  itself)
 can be also expanded in powers of  $\Delta$:
\begin{equation}
\rho(x_1x_2)\equiv\sum_{k=0}^\infty \Delta^k Q_k(x_1x_2),
\end{equation}
Eq.(18) allows one to express the functions  $Q_n$ in terms of $Q_{m}, m<n$. To do this
we substitute the expansions Eq.(62) and Eq.(63) into Eq.(18). We get
\begin{equation}
x_1^2x_2^2\sum_{k=0}^\infty \Delta^k Q_k(x_1x_2)=
\sum_{n,m,k=0}^\infty
{(-1)^{n+m}\Delta^{n+m+k}M_{n+m}\over n!
 m!}\hskip1mm{\partial^{n+m}\over\partial\theta_1^n\partial\theta_2^m}
  \hskip1mmQ_k(\theta_1\theta_2)\bigg|_{\theta_i=U_i-1/x_i}
\end{equation}
 Equating the coefficients at $\Delta^0$ we obtain:
\begin{equation}
x_1^2x_2^2Q_0(x_1x_2)=Q_0(U_1-1/x_1,U_2-1/x_2)
\end{equation}
And consequently (see Eq.(29)) we obtain the following expression for  $Q_0(x_1x_2)$
\begin{equation}
Q_0(x_1x_2)=\rho_0(x_1x_2)={\cal L}_{U_1}(x_1){\cal L}_{U_2}(x_2)
\end{equation}
It is easy to see that if  $M_1=0$ then  $Q_1(x_1x_2)=0$.
 Equating the coefficients at $\Delta^2$ we get:
%\begin{equation}
%x_1^2x_2^2Q_2(x_1x_2)=Q_2(U_1-1/x_1,U_2-1/x_2)+M_2{\partial^2\over\partial\theta_1\partial\theta_2}
%Q_0(\theta_1\theta_2)\bigg|_{\theta_i=U_i-1/x_i}
%\end{equation}
\begin{equation}
x_1^2x_2^2Q_2(x_1x_2)=Q_2(U_1-1/x_1,U_2-1/x_2)+M_2\bigg[{\partial^2\over\partial\theta_1\partial\theta_2}
+{1\over 2}{\partial^2\over\partial\theta_1^2}
+{1\over 2}{\partial^2\over\partial\theta_2^2}\bigg]
Q_0(\theta_1\theta_2)
\bigg|_{\theta_i=U_i-1/x_i}
\end{equation}

%\newpage
 Let us now calculate  up to the terms of $\sim\Delta^3$
 the value of $\langle y_2x_1\rangle$ defined by Eq.(21) with
 $P(\varepsilon)=P_\Delta(\varepsilon)$.
 We use the expansions Eq.(62) and Eq.(63) for the relevant integral:
%\begin{equation}
%\int d\varepsilon P_\Delta(\varepsilon)
%\rho(U_1-\varepsilon+x,U_2-\varepsilon)=
%\int d\varepsilon P_{\Delta}(\varepsilon)\sum_{n,m,k,=0}^\infty{(-\varepsilon)^{n+m}\Delta^k\over n! m!}
%{\partial^{n+m}Q_k(\theta_1\theta_2)\over\partial\theta_1^n\partial\theta_2^m}
%\bigg|_{\theta_1=U_1+x}^{\theta_2=U_2}=
%\end{equation}
%$$
%=\sum_{n,m,k=0}^\infty{(-1)^{n+m}\Delta^{n+m+k}M_{n+m}\over n!m!}
%{\partial^{n+m}Q_k(\theta_1\theta_2)\over\partial\theta_1^n\partial\theta_2^m}
%\bigg|_{\theta_1=U_1+x}^{\theta_2=U_2}=
%$$
%$$
%=Q_0(U_1+x,U_2)+\Delta^2\bigg[Q_2(U_1+x,U_2)+M_2{\partial^2Q_0(\theta_1\theta_2)
%\over\partial\theta_1\partial\theta_2}
%\bigg|_{\theta_1=U_1+x}^{\theta_2=U_2}
%\bigg]+O(\Delta^3)
%$$

\begin{equation}
\int d\varepsilon P_\Delta(\varepsilon)
\rho(U_1-\varepsilon+x,U_2-\varepsilon)=
\int d\varepsilon P_{\Delta}(\varepsilon)\sum_{n,m,k,=0}^\infty{(-\varepsilon)^{n+m}\Delta^k\over n! m!}
{\partial^{n+m}Q_k(\theta_1\theta_2)\over\partial\theta_1^n\partial\theta_2^m}
\bigg|_{\theta_1=U_1+x}^{\theta_2=U_2}=
\end{equation}
$$
=\sum_{n,m,k=0}^\infty{(-1)^{n+m}\Delta^{n+m+k}M_{n+m}\over n!m!}
{\partial^{n+m}Q_k(\theta_1\theta_2)\over\partial\theta_1^n\partial\theta_2^m}
\bigg|_{\theta_1=U_1+x}^{\theta_2=U_2}=Q_0(U_1+x,U_2)+
$$
$$
\Delta^2\bigg[Q_2(U_1+x,U_2)+M_2\bigg({\partial^2
\over\partial\theta_1\partial\theta_2}
+{1\over 2}{\partial^2
\over\partial\theta_1^2}
+{1\over 2}{\partial^2
\over\partial\theta_2^2}\bigg)
Q_0(\theta_1\theta_2)\bigg|_{\theta_1=U_1+x}^{\theta_2=U_2}
\bigg]+O(\Delta^3)
$$

The first term of zero order (with respect to  $\Delta$) is correspond to the
chain with no disorder and therefore do not contribute to the value of $D$ we
are interested in.
 To calculate the  contribution of the first term in square brackets  to $D$
 (we call it $A$-term) one should obtain the function $Q_2(x_1x_2)$. This function can
 be found from Eq.(67).
 Solution of this equation for $|U_{1,2}|<2$ can be performed in the way
similar to that for the  Eq.(26) and reduced to the following redefinition of quantities $J_n(U)$:
\begin{equation}
J_n(U)\equiv\int{dx\over x^2 G^n(x)}
{\partial{\cal L}_U(\theta)\over \partial\theta}\bigg|_{\theta=U-1/x}
\hskip10mm J_n(U)=J_{-n}^\ast(U)
\end{equation}
 By making the following replacing of variable in these integrals: $\theta=U-1/x$
 and making use of the following property of the function $G(\theta)$ Eq.(32):
 $G(1/(U-\theta))=G(\theta)/\lambda_1$\cite{Koz} we get:
\begin{equation}
J_n(U)=\lambda_n\int {d{\cal L}_U(x)\over dx}\hskip1mm {dx\over G^n(x)}
\end{equation}
 Calculation of these integrals shows that  $J_{\pm 1}(U)$ are the only nonzero
 of them:
\begin{equation}
J_{\pm 1}(U)=\mp{\imath\lambda_1\over \sqrt{4-U^2}},\hskip10mm
|J_{\pm 1}(U)|^2={1\over 4-U^2}
\end{equation}
Having this in mind it easy to see that $A$-term from Eq.(68) (when
$|U_{1,2}|<2$) can be written as:
\begin{equation}
\Delta_{\langle y_2x_1\rangle}(t\rightarrow\infty)\bigg|_{\hbox{$A$-term}}=
{\Delta^2 M_2\over 16\pi}\int_{-2}^{2}\sqrt{4-U^2}dU={\Delta^2 M_2\over 8}
\end{equation}
 In the case of  $U_{1,2}>2$ the analysis of Eq.(67) similar to that
 described in the Appendix shows that $A$-term is equal to zero.
  Consider now the contribution of the second term in square brackets of Eq.(68)
  to $D$ (we call it $B$-term).
 When calculating this contribution the region
$U_{1,2}<2$ do not play any role for  the ultimate
(at  $t\rightarrow\infty$) behaviour of  $\Delta_{\langle y_2x_1\rangle}$
 because of the absence of  peculiarities at  $U_1=U_2$ of the functions under
 integrals.
 For  $|U_{1,2}|>2$  $B$-term from Eq.(68) can be written as:
\begin{equation}
\Delta_{\langle y_2x_1\rangle}\bigg|_{\hbox{$B$-term}}=
 M_2\Delta^2\pi\int{dx\over x}{d\over dx}\delta\bigg(
x+{U_1+\hbox{sign}(U_1)\sqrt{U_1^2-4}\over 2}\bigg)\times
\end{equation}
$$
\times{d\over dy}\delta\bigg(
y+{U_2+\hbox{sign}(U_2)\sqrt{U_2^2-4}\over 2}\bigg)
\exp\imath(U_1-U_2)t\bigg|^{t\rightarrow\infty}_{y=0}+...
$$
(Here we present only the terms with crossing derivatives. The remaining terms
can  be analyzed in the same way.)
  Due to the fact that the argument of the second $\delta$-function never becomes
 zero (for  $|U_2|>2$) we come to the conclusion that the $B$-term is equal to
 zero and after multiplying the result Eq.(72) by factor of four we obtain
formulas Eq.(6).

\section{Numerical experiment}

The formulas Eq.(4 -- 6) can be verified by calculation of
 quantities $D$ and function  $W(U)$ by formulas Eq.(2) and Eq.(3)
  where the eigen vectors  ${\bf\Psi}^\lambda$ and the
 eigen energies  $E_\lambda$ are obtained by  direct computer
 diagonalisation of Hamiltonian Eq.(1).
 Below we present the results of such verification.
 The noisy curves were obtained numerically and smooth ones
  by formulas Eq.(4 -- 6).
 Fig.1a shows the dependences of quantity  $D/c$ on the energy of defect
  $\varepsilon$ obtained numerically for different concentrations of defects $c$
   and the relevant theoretical curve Eq.(4).
 It is seen that when  $D/c$ becomes independent on $c$ the Eq.(4)
 is in complete agreement with the numerical results  and the curving
   at $\varepsilon=1$  related to  the non analytical part of Eq.(4)
    is well pronounced.
 Fig.1b shows the dependence of $D$ on the degree
 of disorder  $\Delta$ for the case of  uniform disorder
  when the atomic level splitting distribution function
   has the form Eq.(58) with $p(x)=\Theta(0.5-|x|)$.
  It is seen from this figure that Eq.(6) is in good agreement with
  numerical results even for rather strong disorder.
 Fig.2 shows the "participation" functions  $W(U)dU$ calculated numerically
by Eq.(3) and by analytical formulas Eq.(5) (binary disorder) and Eq.(6)
(uniform disorder).
 Fig.2(a,b)  relate to the case of binary disordered system with
 concentration of defects  $c=0.03$ and energy of defects $\varepsilon=0.8$  (Fig. 2a) and
  $\varepsilon=1.2$ (Fig.2b).
  It is seen that for  $\varepsilon>1$
 the "participation" function " $W(U)$ demonstrate the sharp maximum at
 $U=U_0$ (Eq.(53)).
 Fig.2c shows the case of rather strong  ($\Delta=0.5$) uniform disorder.
 It is seen that in this case  a good agreement between the numerical
 experiment and the theory is also take place but for the description of noticeable dip in the
 center of "experimental" curve one should take into account the corrections of
 higher order than $\Delta^2$.

 No fitting was performed.

\section{Appendix}

To clear up the role of the second item (depending on  $\rho_1$ )
 in Eq.(28) for  $|U_{1,2}|>2$ one should obtain the solution of Eq.(26) for
  this spectral region.
 For the sake of certainty let us  consider the case of $U_{1,2}>2$ and
 introduce the following quantities:
\begin{equation}
 \gamma_U\equiv
{U-\sqrt{U^2-4}\over 2},\hskip5mm
 \gamma'_U=\gamma'_U(\varepsilon)\equiv{1\over U-\gamma_U-\varepsilon},
\end{equation}
where  $\gamma_U$ is the EGF of the chain with no disorder.
 For the case under consideration ( $U_{1,2}>2$) the solution of Eq.(25) gives
the following expression for the function  $\rho_0$:
\begin{equation}
\rho_0(x_1x_2)=\delta(x_1-\gamma_{U_1})\delta(x_2-\gamma_{U_2})
\end{equation}
 Then Eq.(26) can be rewritten as:
\begin{equation}
\rho_1(x_1x_2)-{\rho_1(U_1-1/x_1,U_2-1/x_2)\over x_1^2x_2^2}=
\delta(x_1-\gamma_{U_1}')\delta(x_2-\gamma_{U_2}')-
\delta(x_1-\gamma_{U_1})\delta(x_2-\gamma_{U_2})
\end{equation}
 The direct substitution shows that the solution of Eq.(76) has the form:
\begin{equation}
\rho_1(x_1x_2)=\lim_{M\rightarrow\infty}
\sum_{i=1}^M\bigg\{
\delta(x_1-\theta_i(U_1))\delta(x_2-\theta_i(U_2))-\delta(x_1-\gamma_{U_1})
\delta(x_2-\gamma_{U_2})\bigg\}
\end{equation}
with quantities $\theta_i(U)$ defined by the following recurrent relations:
\begin{equation}
\theta_1(U)=\gamma_{U}',\hskip10mm\theta_{n+1}(U)={1\over U-\theta_n(U)},
 \hskip3mm n=1,2,...,M-1
\end{equation}
 Using Eq.(76) one can write the following expression
  for the function  $\rho_1(U_1+x,U_2)$ entering Eq.(28):
 \begin{equation}
 \rho_1(U_1+x,U_2)=\lim_{y\rightarrow\infty}\bigg({y\over x}\bigg)^2\bigg[
 \rho_1(-1/x,y)-\delta(1/x+\gamma'_{U_1})\delta(y-\gamma'_{U_2})+
\delta(1/x+\gamma_{U_1})\delta(y-\gamma_{U_2})\bigg]
 \end{equation}
 It is clear that the last two terms with $\delta$-functions have zero limit at
 $y\rightarrow\infty$.
  By substituting the function $\rho_1$ (Eq.(77)) to this expression
  one can see that the limit of the first term is also equal to zero.
  Thus, the contribution of the second item in Eq.(28) is equal to zero for
  $|U_{1,2}|>2$.

\newpage

\section*{Captures}

Fig.1 (a): The case of binary disorder.
 Noisy curves -- the dependences of  $D/c$ on the energy of defects  $\varepsilon$
 obtained  numerically for different concentration of defects  $c=0.01, 0.03, 0.1$,
 smooth curve -- the relevant theoretical curve.

 (b): The case of uniform disorder. The dependence of the ultimate at
  $t\rightarrow\infty$ density of excitation  $D$ on the edge site on the
  degree of disorder  $\Delta$.

Fig.2 The "participation" function -- the comparison of theory
(smooth curves)
 and computer simulation (noisy curves).
 For the case of binary disordered chain the occurrence of the peculiarity related to the edge state
 is seen: (a) -- $\varepsilon=0.8<1$, no strong peculiarity,
(b) -- $\varepsilon=1.2>1$, sharp peak appear.
 (c) -- the "participation" function for the chain with  uniform disorder at $\Delta=0.5 $.
 $dU=1/50$ for all cases.

\begin{figure}
\epsfxsize=400pt
\epsffile{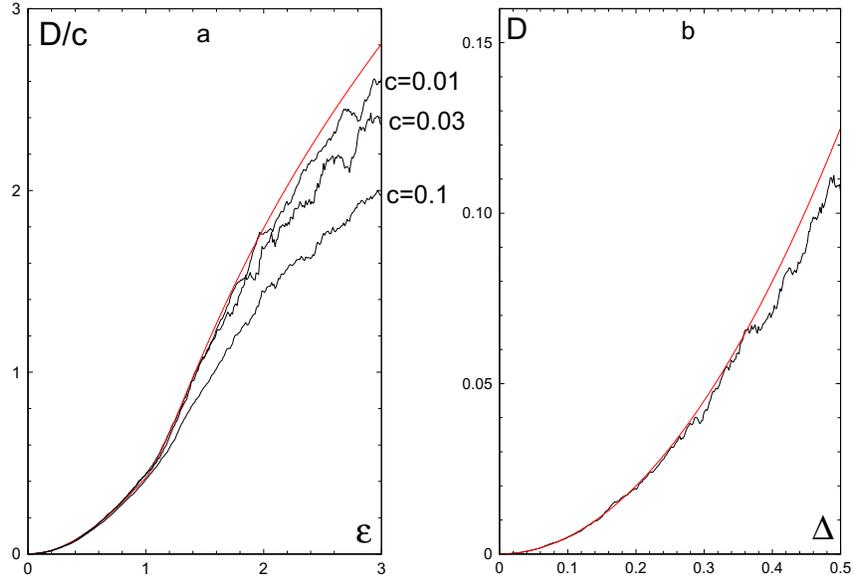}
\caption{(a): The case of binary disorder.
 Noisy curves -- the dependences of  $D/c$ on the energy of defects  $\varepsilon$
 obtained  numerically for different concentration of defects  $c=0.01, 0.03, 0.1$,
 smooth curve -- the relevant theoretical curve.
 (b): The case of uniform disorder. The dependence of the ultimate at
  $t\rightarrow\infty$ density of excitation  $D$ on the edge site on the
  degree of disorder  $\Delta$.}
\end{figure}

\begin{figure}
\epsfxsize=400pt
\epsffile{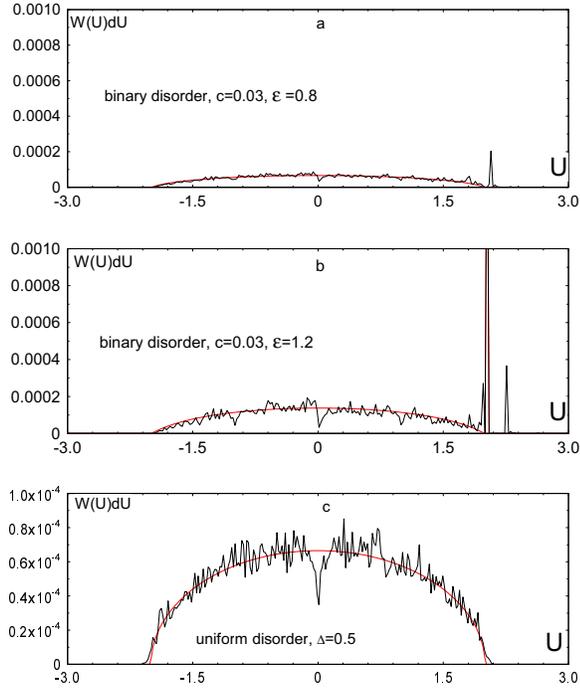}
\caption{The "participation" function -- the comparison of theory (smooth curves)
 and computer simulation (noisy curves).
 For the case of binary disordered chain the occurrence of the peculiarity related to the edge state
 is seen: (a) -- $\varepsilon=0.8<1$, no strong peculiarity,
(b) -- $\varepsilon=1.2>1$, sharp peak appear.
 (c) -- the "participation" function for the chain with  uniform disorder at $\Delta=0.5 $.
 $dU=1/50$ for all cases.}
\end{figure}

\end{document}